INFERENCE WITH IMPUTED DATA: The Allure of Making Stuff Up

Charles F. Manski
Department of Economics and Institute for Policy Research
Northwestern University

May 2022

Abstract

Incomplete observability of data generates an identification problem. There is no panacea for missing data. What one can learn about a population parameter depends on the assumptions one finds credible to maintain. The credibility of assumptions varies with the empirical setting. No specific assumptions can provide a realistic general solution to the problem of inference with missing data. Yet Rubin has promoted random multiple imputation (RMI) as a general way to deal with missing values in public-use data. This recommendation has been influential to empirical researchers who seek a simple fix to the nuisance of missing data. This paper adds to my earlier critiques of imputation. It provides a transparent assessment of the mix of Bayesian and frequentist thinking used by Rubin to argue for RMI. It evaluates random imputation to replace missing outcome or covariate data when the objective is to learn a conditional expectation. It considers steps that might help combat the allure of making stuff up.

1. Introduction

A classic concern of statistics is to use sample data to infer features of a conditional probability distribution. Consider a population characterized by a joint distribution $P(y, x)$, where $y$ is a real outcome and $x$ is a covariate vector. The objective is to learn about $P(y|x)$. One observes $(y_i, x_i, i = 1, . . , N)$ in a random sample of $N$ persons drawn from a study population that has distribution $P(y, x)$. One uses the sample data to estimate features of $P(y|x)$, often the conditional mean $E(y|x)$.

It is well-understood that incomplete observability of sample data generates an identification problem. Inference without assumptions requires contemplation of all logically possible distributions of the missing data. Doing so yields the set of all possible values of $P(y|x)$, its identification region. The practical challenge is to characterize this set in a tractable way. Manski (1989, 1994) showed that identification analysis for $E(y|x)$ and conditional quantiles is simple when only outcome data are missing. Analysis is more complex when the objective is to learn a spread parameter such as $V(y|x)$; see Blundell et al. (2007) and Stoye (2010). Analysis is also more complex when some sample members have missing covariate data. Horowitz and Manski (1998, 2000) study these settings, with focus on $E(y|x)$.

It is also well-understood that assumptions about the distribution of missing data have identifying power. Relatively weak assumptions shrink the identification region for $P(y|x)$. Sufficiently strong assumptions yield point identification. Manski (2003) assembles findings on identification using a spectrum of assumptions.

A basic lesson is that there is no panacea for missing data. What one can learn about a population parameter depends on the assumptions one finds credible to maintain. The credibility of assumptions varies with the empirical setting under study. No specific assumptions can provide a realistic general solution to the problem of inference with missing data.

Yet Rubin (1987, 1996) has promoted random multiple imputation (RMI) as a general, point-identifying, way to deal with missing values in public-use data. The term "imputation" means using artificially constructed values, sometimes called "synthetic data," to take the place of missing data The adjective "random" refers to drawing imputed values at random from a specified probability distribution. The adjective "multiple" refers to repetition of the random imputation process, generating multiple pseudo datasets and correspondingly multiple estimates of parameters of interest. Rubin (1996) made this broad recommendation (p. 473): "For the context for which it was envisioned, with database constructors and ultimate users as distinct entities, I firmly believe that multiple imputation is the method of choice for addressing problems due to missing values."

This recommendation has been influential. Considering missing data in clinical trials, a National Research Council panel (National Research Council, 2010) argued favorably for RMI. Use of RMI has also been recommended for observational studies in medicine (e.g., Sterne *et al.*, 2009; Azur *et al.,* 2011; Pedersen *et al.*, 2017).

Enthusiasm for RMI has extended beyond application to missing data to encompass its use to replace observed data that are deemed sensitive. Rubin (1993) proposed that a data steward use observed data to estimate a model approximating the probability distribution of sensitive data conditional on non-sensitive data. One then uses random draws from this modelled distribution to replace the sensitive data. Repeating this process yields multiple imputations. The idea has subsequently been discussed often. See, for example, Reiter (2002) and Drechsler and Reiter (2009).

It is easy to see why imputation may be attractive to empirical researchers who seek a simple fix to the nuisance of missing data. One constructs synthetic data and then performs statistical analysis as if they are actual data. Rubin (1996) put it this way (p. 474): "Each tool in the ultimate users' existing arsenals can be applied to any data set with missing values using the same command structure and output standards as if there were no missing data." This is the allure of making stuff up.

However, the allure is superficial. RMI is an operational procedure to approximate a modelled distribution of missing by repeated simulation and to facilitate use of readily available statistical software. Performing statistical analysis as recommended by Rubin (1996) has no desirable inferential properties per se. It is well-grounded only if the modelled distribution closely approximates the actual distribution of missing data.

There is rarely good reason to think that modelled distributions of missing data are accurate. It has been common to assume that data are missing at random and to use observed data to model the distribution of missing data. But analysts typically invoke this assumption without much if any comment about its credibility. A remarkably frank exception is a U.S. Census Bureau document describing the American Housing Survey, which states (U. S. Census Bureau, 2011):

> "Some people refuse the interview or do not know the answers. When the entire interview is missing, other similar interviews represent the missing ones . . . . For most missing answers, an answer from a similar household is copied. The Census Bureau does not know how close the imputed values are to the actual values."

Commentators have expressed concern about the accuracy of modelled distributions for sensitive data. Matthews and Harel (2011) observe that inferences made with synthetic data are generically incorrect if the imputation model is incorrect. Reiter (2002) states (p. 532): "the validity of inferences depends critically on the accuracy of the imputation model." He explains that "When these models fail to reflect accurately certain relationships, analysts' inferences also will not reflect those relationships. Similarly, incorrect distributional assumptions built into the models will be passed on to the users' analyses."

I find it disturbing that empirical researchers continue to trust in imputation in general, and RMI in particular, even though maintained assumptions about the distribution of missing data commonly lack foundation. The simplicity of imputation to the user, with its superficial enablement of conventional inference, does not suffice to justify its use in empirical research. Credibility should matter. Rubin's recommendation of RMI as "the method of choice for addressing problems due to missing values" promotes harmful performance of research with incredible certitude (Manski, 2011, 2020).

I have previously cautioned against poorly motivated imputation, notably in Horowitz and Manski (1998) and Manski (2016). Credible assumptions about the distribution of missing data are commonly too weak to yield point identification. Given this, I have recommended that empirical researchers acknowledge the realism of partial identification and report appropriate interval estimates. When point estimation is necessary as an input to decision making, I have recommended that statistical decision theory be used to derive estimates with desired properties (Dominitz and Manski, 2017).

This paper adds to my earlier critiques in three ways. Section 2 aims to provide a transparent assessment of the mix of Bayesian and frequentist thinking used by Rubin to argue for RMI. Section 3, building on and adding to Horowitz and Manski (1998), evaluates random imputation to replace missing outcome or covariate data when the objective is to learn a conditional expectation. Section 4 considers steps that might help combat the allure of making stuff up.

## 2. Rubin's Bayesian Theory and Frequentist Interpretation of RMI

### 2.1. Bayesian Theory

Rubin originally motivated RMI from a subjective Bayesian perspective. One places a joint subjective distribution on all observed and missing data. One computes the posterior subjective distribution of missing data conditional on the observed data. One uses this to derive the posterior distribution of a parameter of interest. Given this, RMI is simply a computational method that uses Monte Carlo integration to approximate the mean of the posterior distribution.

Rubin developed the Bayesian theory in a series of articles and a book (Rubin, 1987). A concise statement was given in Rubin (1996), where he considered the posterior distribution for a real parameter $Q(Y)$, $Y$ being a random vector with some components observed and some missing. He wrote (p. 476):

"The key Bayesian motivation for multiple imputation is given by result 3.1 in Rubin (1987). . . . . the results and its consequences can be easily stated using the simplified notation that the complete-data are $Y = (Y_{obs}, Y_{mis})$, where $Y_{obs}$ is observed and $Y_{mis}$ is missing. Specifically, the basic result is

$$P(Q|Y_{obs}) = \int P(Q|Y_{obs}, Y_{mis})P(Y_{mis}|Y_{obs})dPY_{mis}."$$

Here $P(Q|Y_{obs})$ is the posterior predictive distribution of Q conditional on $Y_{obs}$, $P(Q|Y_{obs}, Y_{mis})$ is the posterior for Q given $(Y_{obs}, Y_{mis})$, and $P(Y_{mis}|Y_{obs})$ is the posterior for $Y_{mis}$ given $Y_{obs}$. Rubin supposed that $P(Q|Y_{obs}, Y_{mis})$ and $P(Y_{mis}|Y_{obs})$ are specified subjective distributions, making $P(Q|Y_{obs})$ computable. In practice, he focused on the posterior mean of Q; that is $E(Q|Y_{obs}) = \int E(Q|Y_{obs}, Y_{mis})P(Y_{mis}|Y_{obs})dPY_{mis}$.

Rubin's "basic result" does not refer to RMI. He interpreted it as RMI by considering Monte Carlo integration as a practical approach to approximate $E(Q|Y_{obs})$. One draws repeated values of $Y_{mis}$ at random from $P(Y_{mis}|Y_{obs})$ and averages the resulting values of $E(Q|Y_{obs}, Y_{mis})$. Semantically, one may refer to Monte Carlo draws of $Y_{mis}$ as imputations. Hence, RMI is Monte Carlo integration.

## 2.2. Frequentist Interpretation

The above motivation for RMI is well-grounded from a subjective Bayesian perspective. A disconnect between the theory and practice of RMI stems from the effort made by Rubin to assert desirable frequentist properties for RMI. To a subjective Bayesian, the posterior mean $E(Q|Y_{obs})$ is interpretable regardless of whether it equals an objective quantity of scientific interest. A frequentist, however, assumes the existence of an objective quantity of interest, say $Q^*$, and wants to estimate this quantity well across repeated samples.

The posterior mean $E(Q|Y_{obs})$ need not be a good estimate of $Q^*$ when $P(Q|Y_{obs}, Y_{mis})$ and $P(Y_{mis}|Y_{obs})$ are simply subjective distributions. To prove good frequentist properties for $E(Q|Y_{obs})$ requires one to assume that the distributions $P(Q|Y_{obs}, Y_{mis})$ and $P(Y_{mis}|Y_{obs})$ are objectively correct. Rubin demonstrated awareness of this core requirement when he wrote (Rubin, 1996, p. 474): "My conclusion is that 'correctly' modeling the missing data must be, in general, the data constructor's responsibility." However, he provided no evidence that data constructors are able to model missing data correctly.

Rubin argued that a desirable frequentist property for statistical procedures is "randomization validity," which he interpreted as requiring approximately unbiased point estimates of scientific estimands. He wrote (Rubin, 1996, p. 476):

"Multiple imputation was designed to satisfy both achievable objectives by using the Bayesian and frequentist paradigms in complementary ways: the Bayesian model based approach to *create* procedures, and the frequentist (randomization-based approach) to *evaluate* procedures."

Continuing, he wrote that if the multiple imputations are "proper" and complete data inference is randomization-valid, then (p. 477): "the large-*m* repeated-imputation inference . . . is randomization-valid for the scientific estimand *Q, no matter how complex the survey design.*"

I find it difficult to understand Rubin's extended verbal discussion of "proper" multiple imputation. However, I believe that I understand the type of frequentist inference that he had in mind. His symbol *m* refers to the number of random draws made from $P(Y_{mis}|Y_{obs})$. Hence, "large-*m*" refers to asymptotic analysis as *m* goes to infinity. Thus, he appears to have meant that Monte Carlo integration yields a well-behaved estimate of a population mean as the number of pseudo-draws goes to infinity. Randomization validity in this sense means that RMI yields a consistent estimate of $E(Q|Y_{obs})$, asymptotically in m. However, it implies nothing about the performance of RMI in estimation of the objective quantity $Q^*$.

### 3. Imputation Estimation of Conditional Expectations

To obtain a concrete sense of the behavior of imputation estimates, this section studies estimation of a conditional expectation when some data are missing and are replaced by imputations. Section 3.1 considers the simple case of imputation of missing outcomes and Section 3.2 the more subtle one of imputation of missing covariates.

### 3.1. Imputation of Missing Outcomes

Consider a population with members characterized by variables (y, x, z). Here y is a real outcome with closed domain Y and x is a covariate vector with finite domain X. Realizations of x are always observable, but some realizations of y are not. The binary variable z indicates whether y is observable (z = 1) or not (z = 0). The distribution of (y, x, z) is denoted P. The objective is to learn $E(y|x = \xi)$ when $P(x = \xi) > 0$.

A simple argument presented in Manski (1989) yields the identification region for $E(y|x = \xi)$. Use the Law of Iterated Expectations to write

(1)  $E(y|x = \xi) = E(y|x = \xi, z = 1)P(z = 1|x = \xi) + E(y|x = \xi, z = 0)P(z = 0|x = \xi).$

$E(y|x = \xi, z = 1)$ and $P(z = 1|x = \xi)$ are observable but $E(y|x = \xi, z = 0)$ is not. Knowledge of the domain Y restricts $E(y|x = \xi, z = 0)$ to lie in the interval $[Y_L, Y_U]$, where $Y_L \equiv \min(Y)$ and $Y_U \equiv \max(Y)$. Hence, the identification region with no assumptions on the distribution of missing data is the interval

(2)  $[E(y|x = \xi, z = 1)P(z = 1|x = \xi) + Y_L P(z = 0|x = \xi),\ E(y|x = \xi, z = 1)P(z = 1|x = \xi) + Y_U P(z = 0|x = \xi)].$

If assumptions restrict $E(y|x = \xi, z = 0)$ to a proper subset of $[Y_L, Y_U]$, say $\Gamma$, the identification region is

(3)  $E(y|x = \xi, z = 1)P(z = 1|x = \xi) + \gamma \cdot P(z = 0|x = \xi), \gamma \in \Gamma.$

Random imputation estimates assume that $P(y|x = \xi, z = 0)$ is a specified distribution and use realizations drawn from this distribution to replace missing values of y. Suppose that a random sample of N population members are drawn. One observes $(x_i, z_i)$ for all $i = 1, \ldots, N$ and observes $y_i$ when $z_i = 1$. If y were always observed, one might naturally estimate $E(y|x = \xi)$ by the sample average $E_N(y|x = \xi,)$. To cope with missing outcome data, consider replacement of missing values of y with imputations and computation of the sample average combining observed and imputed data.

I examine the probability limit of the imputation estimate as sample size goes to infinity, showing how the limit depends on the probability distribution used to generate imputations. It suffices to study the limiting behavior of estimation using one synthetic sample. Multiple imputation yields multiple synthetic samples and multiple estimates, each with the same probability limit.

### 3.1.1. Analysis

Let each member of the population be assigned an imputed outcome u ∈ Y, which is used to replace missing outcome data. In the sample of size N, Let N(1, ξ) be the sub-sample of cases where (z = 1, x = ξ). Let N(0, ξ) be the sub-sample where (z = 0, x = ξ). Let $N_{1\xi} = |N(1, \xi)|$, $N_{0\xi\omega} = |N(0, \xi)|$, and $\pi_{N\xi} \equiv N_{1\xi}/(N_{1\xi} + N_{0\xi})$. Whenever $N_{1\xi} + N_{0\xi} > 0$, the imputation estimate of E(y|x = ξ) is

$$(4) \quad \theta_{N\xi} \equiv \frac{1}{N_{1\xi} + N_{0\xi}} \left( \sum_{i \in N(1, \xi)} y_i + \sum_{i \in N(0, \xi)} u_i \right)$$

$$= \pi_{N\xi} \frac{1}{N_{1\xi}} \sum_{i \in N(1, \xi)} y_i + (1 - \pi_{N\xi}) \frac{1}{N_{0\xi}} \sum_{i \in N(0, \xi)} u_i \, .$$

Let N → ∞. The probability limit of $\theta_{N\xi}$ is

$$(5) \quad \theta_\xi \equiv E(y|x = \xi, z = 1) \cdot P(z = 1|x = \xi) + E(u|x = \xi, z = 0) \cdot P(z = 0|x = \xi).$$

In general, $\theta_\xi \neq E(y| x = \xi)$. Comparison of (5) with (1) shows that the imputation estimate is consistent if and only if

$$(6) \quad E(u|x = \xi, z = 0) = E(y|x = \xi, z = 0).$$

Seeking to justify (6), researchers sometimes assume that data are missing at random conditional on x and aim to draw imputations from a consistent estimate of the observable distribution P(y|x, z = 1), say $P_N(y|x, z = 1)$. Thus, let

$$(7a) \quad P(y|x = \xi, z = 0) = P(y|x = \xi, z = 1),$$

(7b)  $P(u|x = \xi, z = 0) = P_N(y|x = \xi, z = 1)$.

Equation (7a) is an untestable assumption. Given (7a), equation (7b) asymptotically implies (6).

### 3.1.2. Alternatives to Imputation

Alternatives to imputation provide superior approaches to inference on $E(y|x)$. First suppose that one thinks it credible to assume that missing data have a particular distribution, say $Q(y|x = \xi, z = 0)$. Let $E_Q(y|x = \xi, z = 0)$ be the mean outcome under Q. Then a simple alternative to imputation is to replace missing values by $E_Q(y|x = \xi, z = 0)$, yielding the estimate

(8)  $\theta_{QN\xi} \equiv \pi_{N\xi} \dfrac{1}{N_{1\xi}} \displaystyle\sum_{i \in N(1, \xi)} y_i + (1 - \pi_{N\xi}) \cdot E_Q(y|x = \xi, z = 0)$.

Estimate (8) has the same probability limit as the imputation estimate (4). Moreover, it has greater statistical precision.

Now suppose that one lacks a credible basis to specify a distribution for missing data and contemplates estimation without assumptions. Interval (2) gives the identification region for $E(y|x = \xi)$. This interval is bounded if $Y_L$ and $Y_U$ are finite. Then a natural interval estimate for $E(y|x = \xi)$ is the sample analog of (2), namely

(9)  $[\pi_{N\xi} \dfrac{1}{N_{1\xi}} \displaystyle\sum_{i \in N(1, \xi)} y_i + (1 - \pi_{N\xi})Y_L, \quad \pi_{N\xi} \dfrac{1}{N_{1\xi}} \displaystyle\sum_{i \in N(1, \xi)} y_i + (1 - \pi_{N\xi})Y_U]$,

whose probability limit is (2).

Should it be necessary to provide an assumption-free point estimate of $E(y|x)$, an attractive option is to use the midpoint of interval (9). This estimate converges to the midpoint of (2), which minimizes the maximum value of asymptotic squared bias among all point estimates of $E(y|x)$. Dominitz and Manski (2017) study the finite-sample performance of the midpoint estimate from the perspective of statistical decision theory and derive the maximum value of mean square error across all distributions of missing data.

## 3.2. Imputation of Missing Covariates

In practice, many patterns of missing data may occur within a covariate vector. Analysis of every possible pattern requires cumbersome notation, so I focus on settings in which some covariates are always observed whereas others may have missing data. I denote the former covariates as x and the latter as w. Thus, consider a population with members characterized by variables $(y, x, w, z)$. Here y is a real outcome with domain Y, whereas x and w are covariate vectors with finite domains X and W. Realizations of $(y, x)$ are always observable, but some realizations of w are not. The binary variable z now indicates whether w is observable $(z = 1)$ or not $(z = 0)$. The population distribution of $(y, x, w, z)$ is P. The objective is to learn $E(y|x = \xi, w = \omega)$ when $P(x = \xi, w = \omega) > 0$.

Horowitz and Manski (1998) derived the identification region for $E(y|x, w)$ with no assumptions on the distribution of missing data. The derivation is much more subtle than with missing outcome data and the general form of the region is much more complex than (2). However, Horowitz and Manski (2000) and Manski (2003) show that the region has a simple explicit form when y is a binary outcome, say taking the values 0 and 1. Applying Manski (2003, Corollary 3.8.1), the identification region for $E(y|x = \xi, w = \omega)$ is then the interval

$$(10) \quad \left[ \frac{P(y = 1, x = \xi, z = 1)}{P(x = \xi, z = 1) + P(y = 0, z = 0)} \; , \; \frac{P(y = 1, x = \xi, z = 1) + P(y = 1, z = 0)}{P(x = \xi, z = 1) + P(y = 1, z = 0)} \right].$$

Random imputation estimates assume that $P(w|y, x = \xi, z = 0)$ is a specified distribution and use realizations drawn from this distribution to replace missing values of w. Suppose that a random sample of N population members are drawn. One observes $(y_i, x_i, z_i)$ for all $i = 1, \ldots, N$ and observes $w_i$ when $z_i = 1$.

If y were always observed, one might naturally estimate $E(y|x = \xi, w = \omega)$ by the sample average $E_N(y|x = \xi, w = \omega)$. To cope with missing covariate data, consider replacement of missing values of w with imputations and computation of the sample average combining observed and imputed data.

I examine the probability limit of the estimate as sample size goes to infinity, showing how the limit depends on the probability distribution used to generate imputations. It again suffices to study the limiting behavior of estimation using one synthetic sample as multiple imputation yields multiple estimates, each with the same probability limit.

### 3.2.1. Analysis

Let each member of the population be assigned an imputed value $u \in W$, which is used to replace missing covariate data. In the sample of size N, Let $N(1, \xi, \omega)$ be the sub-sample of cases where $(z = 1, x = \xi, w = \omega)$. Let $N_m(0, \xi, \omega)$ be the sub-sample where $(z = 0, x = \xi, u_m = \omega)$. Let $N_{1\xi\omega} = |N(1, \xi, \omega)|$, $N_{0\xi\omega} = |N(0, \xi, \omega)|$, and $\pi_{N\xi\omega} \equiv N_{1\xi\omega}/(N_{1\xi\omega} + N_{0\xi\omega})$. Then, whenever $N_{1\xi\omega} + N_{0\xi\omega} > 0$, the imputation estimate of $E(y|x = \xi, w = \omega)$ is

$$(11) \quad \theta_{N\xi\omega} \equiv \frac{1}{N_{1\xi\omega} + N_{0\xi\omega}} \left( \sum_{i \in N(1, \xi, \omega)} y_i + \sum_{i \in N(0, \xi, \omega)} y_i \right)$$

$$= \pi_{N\xi\omega} \frac{1}{N_{1\xi\omega}} \sum_{i \in N(1, \xi, \omega)} y_i + (1 - \pi_{N\xi\omega}) \frac{1}{N_{0\xi\omega}} \sum_{i \in N(0, \xi, \omega)} y_i .$$

Let $N \rightarrow \infty$. The probability limit of $\theta_{N\xi\omega}$ is

$$(12) \quad \theta_{\xi\omega} \equiv E(y|x = \xi, w = \omega, z = 1) \cdot \pi_{\xi\omega} + E(y|x = \xi, u = \omega, z = 0) \cdot (1 - \pi_{\xi\omega}),$$

where

$$(13) \quad \pi_{\xi\omega} = \frac{P(z = 1, x = \xi, w = \omega)}{P(z = 1, x = \xi, w = \omega) + P(z = 0, x = \xi, u = \omega)}$$

$$= \frac{P(z = 1, w = \omega \mid x = \xi)}{P(z = 1, w = \omega \mid x = \xi) + P(z = 0, u = \omega \mid x = \xi)} \ .$$

In general, $\theta_{\xi\omega} \neq E(y \mid x = \xi, w = \omega)$. By the Law of Iterated Expectations,

$$(14) \quad E(y \mid x = \xi, w = \omega) = E(y \mid x = \xi, w = \omega, z = 1) \cdot P(z = 1 \mid x = \xi, w = \omega)$$

$$+ E(y \mid x = \xi, w = \omega, z = 0) \cdot P(z = 0 \mid x = \xi, w = \omega).$$

Comparison of (12) and (14) shows that they coincide if

$$(15a) \quad P(z = 0, u = \omega \mid x = \xi) = P(z = 0, w = \omega \mid x = \xi),$$

$$(15b) \quad E(y \mid x = \xi, u = \omega, z = 0) = E(y \mid x = \xi, w = \omega, z = 0).$$

These equalities generically do not hold.

Equations (15a)-(15b) do hold if the distribution of imputations is

$$(16) \quad P(u \mid y, x, z = 0) = P(w \mid y, x, z = 0).$$

Multiplying both sides of (16) by the observable distribution $P(y, x, z = 0)$ yields

$$(17) \quad P(y, x, u, z = 0) = P(y, x, w, z = 0),$$

which implies (15a)-(15b). The problem, of course, is that satisfaction of equation (16) requires knowledge of the distribution $P(w|y, x, z = 0)$ of missing data.

Seeking to justify (16), researchers sometimes assume that data are missing at random conditional on $(y, x)$ and aim to draw imputations from a consistent estimate of the observable distribution $P(w|y, x, z = 1)$, say $P_N(w|y, x, z = 1)$. Thus, let

(18a)  $P(w|y, x, z = 0) = P(w|y, x, z = 1)$,

(18b)  $P(u|y, x, z = 0) = P_N(w|y, x, z = 1)$.

Equation (18a) is an untestable assumption. Given (18a), equation (18b) asymptotically implies (16).

### 3.2.2. Imputation as an Attempted Solution to the Ecological Inference Problem

A polar case of missing covariates that warrants separate attention occurs when a sample of $(y, x, w)$ is drawn but the realizations of w are never observed; thus, $P(z = 0) = 1$. No conclusions about $E(y| x = \xi, w = \omega)$ can be drawn without further information. In some settings, a separate sampling process yields observations of $(w, x)$ but not of y. Then one faces the problem of ecological inference; that is, inference on the "long" conditional distribution $P(y|x, w)$ given observability of the "short" distributions $P(y|x)$ and $P(w|x)$.

Duncan and Davis (1953) considered identification of $P(y = 1| x = \xi, w = \omega)$ when y is a binary outcome. They sketched a proof that, in the absence of assumptions restricting the long distribution, the identification region is the interval

(19)    $[0, 1] \cap [P(y = 1|x = \xi) - P(w \neq \omega|x = \xi)]/P(w = \omega|x = \xi), P(y = 1|x = \xi)/P(w = \omega|x = \xi)]$.

Horowitz and Manski (1995) formalized this finding and studied identification of E(y| x = ξ, w = ω) when y is real-valued. The analysis is much more subtle than when y is binary. The identification region is an interval that does not have an explicit form but can be computed numerically.

Some medical researchers have used imputation of genotypes in an attempt to solve the ecological inference problem. In this setting, y is a patient outcome while (x, w) are genetic markers. One dataset yields observations of (y, x) and another provides observations of (x, w), enabling assumption-free estimation of P(y|x) and P(w|x). For each patient i in the former dataset, the estimate of distribution P(w|x = $x_i$) is used to impute $w_i$, yielding ($y_i$, $x_i$, $u_i$), i = 1, . . . , N. This partially synthetic dataset is analyzed using the imputations as if they were real covariate data. See Gragert *et al.* (2014), Tinckam *et al.* (2016), Geneugelijk *et al.(*2017), Kamoun *et al.* (2017), and Nilsson *et al.* (2019).

Manski, Tambur, and Gmeiner (2019) and Manski, Gmeiner, and Tambur (2021) counsel against this use of genotype imputation. By construction, imputations drawn from P(w|x) are statistically independent of actual patient outcomes y. Applying (12), the probability limit of the imputation estimate of the long conditional mean E(y| x = ξ, w = ω) is the short conditional mean E(y| x = ξ). Thus, imputation of w accomplishes nothing.

### 3.2.3. Alternatives to Imputation

Suppose one thinks it credible to assume that missing w data have a particular distribution, say Q(w|y, x = ξ, z = 0). This assumption can be used to estimate E(y|x, w) without constructing imputed data. Use the Law of Total Probability to write

(19)    P(y, x, w)  =  P(y, x, w|z = 1)P(z = 1) + P(y, x, w|z = 0)P(z = 0)

=  P(y, x, w|z = 1)P(z = 1) + P(w|y, x, z = 0)P(y, x|z = 0P(z = 0).

Distributions P(y, x, w|z = 1), P(y, x|z = 0), and P(z) are observable. Each is consistently estimable by its sample analog, denoted $P_N$(y, x, w|z = 1), $P_N$(y, x|z = 0), and $P_N$(z). Inserting these estimates into (19), and assuming that Q is the distribution of missing covariates, yields a consistent estimate of P(y, x, w), namely

(20)   $P_N$(y, x, w) = $P_N$(y, x, w|z = 1)$P_N$(z = 1) + Q(w|y, x, z = 0)$P_N$(y, x|z = 0)$P_N$(z = 0).

The conditional mean of $P_N$(y, x, w) is a consistent estimate of E(y|x, w).

Suppose that one lacks a credible basis to specify a distribution for missing data and contemplates estimation without assumptions. The complex general form of the identification region for E(y|x = ξ, w = ω) makes estimation of this region complex as well. However, estimation is easy when y is a binary outcome. Then interval (10) gives the identification region. A natural interval estimate is the sample analog of (10), namely

(22)   $\left[ \dfrac{P_N(y = 1, x = \xi, z = 1)}{P_N(x = \xi, z = 1) + P_N(y = 0, z = 0)} \ , \ \dfrac{P_N(y = 1, x = \xi, z = 1) + P_N(y = 1, z = 0)}{P_N(x = \xi, z = 1) + P_N(y = 1, z = 0)} \right]$,

whose probability limit is (10). Should it be necessary to provide an assumption-free point estimate, an attractive option is to use the midpoint of interval (22). As N increases this converges to the midpoint of (10), which minimizes the maximum value of asymptotic squared bias among all point estimates.

## 4. Combatting the Allure of Making Stuff Up

The use of imputed data is a striking illustration of research with incredible certitude. Arguing for RMI, Rubin (1996) wrote (p. 473): "alternative methods either require special knowledge and techniques not available to typical users or produce answers that are generally not statistically valid for scientific estimands." Development of user-friendly methods is a worthy objective, provided that the methods yield useful findings.

The findings of RMI are useful only if the assumed distribution of missing data is close to correct. Rubin recognized this central requirement in principle when he wrote (p. 474): "My conclusion is that 'correctly' modeling the missing data must be, in general, the data constructor's responsibility." Yet assigning responsibility to the data constructor is futile if demonstrably correct modeling is not achievable. Assumed distributions of missing data commonly lack credible foundations. Hence, Rubin's assertions that RMI yields findings that are valid in certain Bayesian and frequentist senses should not comfort empirical researchers who want to make credible inferences about the real world.

Clarification of the fragility of imputation is necessary to combat the allure of making stuff up, but I doubt that it will suffice. Another necessary step is to provide tractable statistical methods that enable credible empirical research. Section 3.1 showed that assumption-free interval estimation is simple with missing outcome data. Section 3.2 showed that it is simple with missing covariate data when y is a binary outcome. Interval estimation bringing to bear some potentially credible assumptions with identifying power is straightforward. Examples include monotone-instrumental-variable and bounded-variation assumptions (Manski and Pepper, 2000; Manski, Tambur, and Gmeiner, 2019). However, identification regions in some problems with missing data are complex, making interval estimation difficult. Empirical researchers should be sophisticated enough to recognize that performing credible analysis is sometimes a challenge.